\documentclass[manuscript]{aastex}

\usepackage{lscape}

\shorttitle{Spectroscopy  of Optically Selected BL Lac Objects. }
\shortauthors{A. Sandrinelli, R. Falomo,  A. Treves.}

\begin{document}


\title{Spectroscopy of Optically Selected BL Lac Objects and their $\gamma$-ray emission 
}

\author{A. Sandrinelli}
\affil{\textit{Universit\`a degli Studi del$\l'$Insubria, Via Valleggio 11, I-22100 Como, Italy\\
INAF - Osservatorio Astronomico di Brera, Via Emilio Bianchi 46, I-23807 Merate, Italy
INFN - Istituto Nazionale di Fisica Nucleare}}
\email{angela.sandrinelli@brera.inaf.it}

\author{A. Treves}
\affil{\textit{Universit\`a degli Studi del$\l'$Insubria, Via Valleggio 11, I-22100 Como, Italy\\
INAF - Istituto Nazionale di Astrofisica\\
INFN - Istituto Nazionale di Fisica Nucleare}}

\author{R. Falomo}
\affil{\textit{INAF - Osservatorio Astronomico di Padova, Vicolo dellÕ Osservatorio 5, I-35122 Padova, Italy}}

\author{E. P. Farina}
\affil{\textit{Universit\`a degli Studi del$\l'$Insubria, Via Valleggio 11, I-22100 Como, Italy
\\INFN - Istituto Nazionale di Fisica Nucleare}}

\author{L. Foschini}
\affil{\textit{
INAF - Osservatorio Astronomico di Brera, Via Emilio Bianchi 46, I-23807 Merate, Italy}}

\author{M. Landoni}
\affil{\textit{Universit\`a degli Studi del$\l'$Insubria, Via Valleggio 11, I-22100 Como, Italy\\
INAF - Osservatorio Astronomico di Brera, Via Emilio Bianchi 46, I-23807 Merate, Italy
INFN - Istituto Nazionale di Fisica Nucleare}}

\and

\author{B. Sbarufatti}
\affil{\textit{
INAF - Osservatorio Astronomico di Brera, Via Emilio Bianchi 46, I-23807 Merate, Italy\\
Pennsylvania State University, Department of Astronomy and Astrophysics, PA 16801, USA}}



\begin{abstract}

We present Very Large Telescope optical spectroscopy\footnote{Based on observations made with ESO Telescopes at the Paranal Observatory under program 083.B-0187(A).} of nine BL Lac objects of unknown redshift belonging to the list of optically selected 
radio loud BL Lacs candidates. We explore their spectroscopic properties and 
the possible link with gamma ray emission. From the new observations we determine the redshift of four objects  from faint emission lines or from absorption features of the host galaxy.
In three cases we find narrow intervening absorptions from which a lower limit to the redshift is inferred.
For the remaining two featureless sources, lower limits to the redshift  are deduced from the very absence of  spectral lines.
A search for $\gamma$ counterpart emission shows that  six out of nine are \textit{Fermi} $\gamma$-ray emitters with two new detections.  Our analysis suggests that most of the BL Lac still lacking of redshift information are most probably located at high redshift. 

\end{abstract}


\section{Introduction}
BL Lac objects are active  galactic nuclei (AGNs) characterized by strong and rapid flux variability, polarization, and weakness or absence of spectral emission lines. With  the flat spectrum radio quasars (FSRQ), BL Lacs represent  a type of radio loud objects called blazars. As proposed in the seminal paper of  \cite{Blandford1978}, blazars are AGNs with relativistic jets pointing close to the direction of the observer. They are the dominant population of the extragalactic  $\gamma$-ray sky both at GeV and TeV energies. 
 At radio frequencies,  BL Lacs display  strong core compact flat spectrum emissions. In the optical range, synchrotron continuum  is boosted by relativistic beaming  resulting in a depression of the equivalent width of the spectral lines, especially in the high states spectra, making often the detection of the  redshift a challenging task and a central issue in $\gamma$ astronomy.

The first complete surveys of BL Lac objects were performed in the radio band  considering as distinguish feature the flatness of the radio spectrum \citep[e.g.][]{Stickel1991}. In X-rays the BL Lac surveys were  a sub-product of "complete" sky surveys \cite[see e.g.][] {Stocke1985,Stocke1991}. The Palomar Green survey \citep{1986ApJS...61..305G}, aimed to build a complete list of quasars through their colors, gave four bright BL Lacs.
   The  optical  spectroscopy  was consequent,   since the commonly used criterion for defining a BL Lac is the line equivalent widths $EW  < 5$ \AA \ \citep[e.g.][]{Marcha1996}.
 The detection of these weak features  requires spectroscopy of adequate spectral resolution and signal to noise. Observations with medium aperture telescopes provided the redshift of a number of BL Lacs \citep[e.g.,][]{Falomo1987b,Falomo1987a,Ulrich1989,Falomo1990,Stickel1993,Veron1994, Marcha1996,Carangelo2003} 
but for many of them, in particular those with a strong nuclear component, the redshift remained unknown. With 8m class telescopes the situation improved as demonstrated, among the recent systemtic spectroscopic campaigns,  by our study  of $69$ BL Lacs  in the southern sky  with ESO-VLT+FORS2, which yielded 23 new redshifts of BL Lacs basically selected in the Giommi/Padovani list \citep{Padovani1995} before the \textit{Fermi} launch \bibpunct[; ]{(}{)}{,}{a}{}{,} \citep [][spectra and redshifts are available in electronic form in our website \texttt{http://www.oapd.inaf.it/zbllac/}]{Sbarufatti2005a,  Sbarufatti2005b, Sbarufatti2006a,  Sbarufatti2006b, Sbarufatti2009, Landoni2013}.

In the last decade in parallel with the activity related to the high energy emission, a substantial progress in discovering new BL Lacs and their redshift derived from large optical spectroscopic surveys, in combination with radio and X-ray catalogs.
\citet[][ hereafter P08]{Plotkin2008} selected  501 BL Lacs candidates combining
 the Faint Images of the Radio Sky at Twenty-Centimeters \citep[FIRST,][]{Becker1995} radio survey
with  the Sloan Digital Sky Survey Data Release 5 (SDSS DR5) spectroscopic data base, under the  conditions of featureless or  weak features spectra and Ca II H/K depression less than $40 \%$. A substantial fraction of sources,  ($\sim60\%$) lacks of reliable spectroscopic redshifts.
Recently,  \citet[][hereafter S13]{Shaw2013} using different telescopes,  produced spectra of most of the \textit{Fermi}  475 BL Lac candidates \citep{Ackermann2011}, obtaining redshifts for $\sim44\%$ of the sample and  constraining $z$ for nearly all remaining objects. 
 However, in order to characterize the general properties of the BL Lac population it is highly desirable to define an homogeneous sample  of BL Lacs   not biased by the properties introduced in the selection by the X -rays and radio surveys. 
For instance, \cite{Collinge2005} compiled  a large  optically selected sample (386 targets) from 2860 deg$^2$  of the SDSS, chosen to have quasi-featureless optical spectra and low proper motions.
Some radio-quiet sources were found, almost all without  X-ray  counterparts in ROSAT All-Sky Survey \citep[RASS;][]{Voges1999}.
\citet[][hereafter P10]{Plotkin2010a} expanded the Collinge sample through a complex sieving procedure of SDSS DR7 \citep{Abazajian2009}, and recovered  723 purely optically selected BL Lacs,  included a fraction of 86 radio-quiet objects, the majority of which are unlikely \textit{bona fide} BL Lacs, but rather a distinct class of quasars with intrinsically weak emission lines \citep{Plotkin2010b,Wu2012}. Approximately $\sim80\%$ of the whole P10 sample match to a radio source in FIRST/NRAO VLA Sky Survey \citep[NVSS;][]{Condon1998}, and $\sim40\%$ match to a RASS X-ray source.
Spectroscopic redshifts are given for $\sim36\%$ of the radio-loud subsample.

For this elusive class of objects the adopted selection criteria can affect the redshift distributions of the BL Lacs
and  cause  different cosmological evolution scenarios, see e.g. the discussions in \cite{Bade1998} and in \cite{Giommi2012}. 
Radio selected BL Lacs seem to display  a positive  evolution (i.e. the number density or the luminosity shows a decrease with cosmic time),
while for X-ray selected objects  a negative one was proposed, or not evolvution at all \bibpunct[]{(}{)}{,}{a}{}{,}\citep[][ and references therein]{Rector2000,Rector2001,Caccianiga2002,Beckmann2003,Padovani2007,Ajello2009,Giommi2009, Giommi2012}.
A continuum trend  from slightly positive-evolution low peaked BL Lacs to strong negative-evolution high peaked was proposed \citep{Rector2000}, also related to their X-ray to radio flux ratio \citep{Giommi1999,Giommi2012}. Statistics  concerning the evolution of the BL Lacs suffer of redshift incompleteness,  making  the increase of the objects  with reliable redshifts from homogeneous and unbiased selections a core issue \citep[see also][ for a discussion]{Shaw2013}.

In this paper we present  optical spectroscopy of a small sample (9 targets) of BL Lacs of still unknown redshift belonging to the P08 catalog of radio selected BL Lacs.  
We note that our sample is also entirely included  into P10 catalog of optically selected BL Lacs. We describe our observations and analysis of spectra in section 2 together with the new redshifts; 
then we search for the counterparts in the \textit{Fermi} Gamma-ray Space Telescope archives (section 3). Summary and conclusions are given in section 4.
\\

Throughout the paper we adopt the following concordant cosmology:  $H_0=70$ $km$  $s^{-1}$  $Mpc^{-1}$, $\Omega_m =0.30$ and $\Omega_\Lambda=0.70$.

\section{Optical Spectroscopy}
\subsection{Sample observations and data analysis}

Within about 200 P08 BL Lacs objects with unknown redshift, we selected a small sample of 15 with  the only requests that they were relatively bright objects ($r<19.4$),  well observable from  Paranal (Chile) ESO premises and classified as high confidence BL Lacs.  We stress that these are also all  included in the P10 radio loud BL Lac  candidates, selected on the only basis of they optical properties, thus we can  consider them  as \textit{a posteriori} optically selected. We collected
 optical spectra of only 9 sources (see {Table \ref{journal}) out of 15  because of weather conditions.

Spectra  were gathered with the FORS2 mounted on the Antu VLT  of the ESO in Paranal.
Observations were performed  with the grism 300V and the
2$^{\prime\prime}$ width slit, yielding a spectral 
resolution at the central wavelength of $R=\lambda/\Delta
\lambda\sim 350$  and covering $3800-8200$ \AA\  spectral range, and exploiting the better S/N ratio of VLT.
The seeing in the nights of observations ranged from $0\farcs5$ to $1\farcs2$ , with an average of $\sim 0\farcs9$, as reported in Table \ref{journal}.  
 Standard IRAF\footnote{\texttt{IRAF} \citep{Tody1986} is distributed by the National Optical Astronomy 
Observatories, which are operated by the Association of 
Universities for Research in Astronomy, Inc., under 
cooperative agreement with the National Science Foundation.} 
tools were used in data reduction.
We adopted the same procedures described in previous works \citep[e.g.][]{Sbarufatti2005a,Sbarufatti2005b}, 
including bias subtraction and flat fielding. For each target we obtained three or six individual spectra to make corrections for the cosmic rays  and provide independent check of each signature, with typical total exposure times of 45 or 90 min respectively (see  Table \ref{journal}). Individual spectral frames  are combined together taking the median, from which  a one-dimensional spectrum is extracted. 
The wavelength calibration was achieved using the spectra of a Helium Neon Argon lamp and  typical uncertainties  are $\sim1$ \AA.
Spectra are corrected for Galactic reddening according with the \citet{Schlegel1998} maps and assuming R$_{\rm V}=3.1$ 
\citep[e.g.][]{Cardelli1989}.

\subsection{The optical spectra}

 The extracted spectra and the normalized spectra with respect to the continuum are reported  both in Figure \ref{4_1}  and in electronic form in our mentioned website.   
For each spectrum the  S/N is given in Table \ref{journal}. 
The continuum was fitted with a power law, defined as $F_{\lambda} \propto \lambda^{-\alpha_{\lambda}}$. The resulting optical spectral indices are given  in the Table \ref{par} 
as $\alpha_{\nu}$ ($F_{\nu} \propto \nu^{-\alpha_{\nu}}$, where $\alpha_{\nu}=2-\alpha_{\lambda}$) for consistency and easy comparison with the bulk of the literature. We find $0.73< \alpha_{\nu}< 1.44$ corresponding to average value  $\alpha_{\nu \ ave}=1.17$,  both consistent with  the average spectral index $\alpha_{\nu}=1.15$ and the dispersion of 0.69 reported by P10.

In order to search for very weak spectral lines, we evaluated the minimum observable equivalent width ($EW_{min}$). 
Dividing the spectrum in 25 \AA \ bins,  as fully described in \cite{Sbarufatti2005b}, we  objectively define the  $EW_{min}$ as twice the rms of the distribution of all the observer-frame $EWs $ measured in each bin. 
The absorption and emission features with $EW$ greater than this threshold are carefully inspected.   
ID labels mark the successful identifications in Figure \ref{4_1} and in  Table  \ref{par}, where  also $EW_{min}$ and line properties are reported.
For four sources  we were able to obtain a redshift from the detection absorption/emission lines associated to the BL Lac host galaxy. In three cases, we observed absorption intervening features that,  interpreted as Mg II  2800, allow us to set a lower limit to the redshift. 
In two cases the spectrum is featureless, thus we calculated a lower limit on $z$ using the method described by \cite{Sbarufatti2005b, Sbarufatti2006a}.
 Briefly, recalling that non thermal jet and the host galaxy both  contribute to the observed flux and assuming  that BL Lac host galaxies are  giant elliptical with $M_R^{host}=-22.90\pm0.50$ and  standard absorption lines, one can 
infer a $z$ dependent relationship between observed $EW$ and emitted $EW_{rest}= EW/(1+z)$ equivalent widths.  
The absence of lines places $EW<EW_{min}$, providing a lower limit for $z$. In Table \ref{par} we also report the redshift  upper limit from the lack  of the $Ly{\alpha}$ absorptions as $z_{upper}=\lambda_{lim} / 1216 $ \AA$ -1+ \Delta z$  , where $\Delta z$ derives from  the probability of not detection absorbers close to the blue limit  wavelength $\lambda_{lim}$ of the spectrum, taking into account the redshift dependence of $Ly{\alpha}$ forest absorbers density and its $EW$ scaling \citep[see][ and references therein]{Shaw2013}.

In the following we report further information on single sources.

 \subsection{Notes on individual sources.}

\textbf{J003808.50+001336.5}:   The spectrum shows a feature at $\lambda=4780$  \AA. We ascribed to Mg II $\lambda2798$ intervening  system absorption, which places  the source at $z > 0.708$.
This feature is,  however,  only detected at $2\sigma$ level. We also note that Mg II doublet is unresolvable with our observations. To complement our observations we retrieved all the spectra from the SDSS. This target was observed 3 times from MJD 51793 to 55444  with  $S/N\sim10$ and no  reliable redshift is available.
\\

\textbf{J125032.58+021632.1}:  Mg II $\lambda2798$ and O II $\lambda3727$ emission lines are apparent, securing the source at $z = 0.995$. A tentative redshift of 0.953, warning flagged for multiple equal  $\chi^2$ best fits, was assigned to the source by the  redshift fitting procedure in the SDSS DR9 based on a  spectrum of $S/N\sim10$, where the Mg II feature is loosely visible and detectable.
For this optically selected BL Lac candidate the rest frame Mg II  equivalent width is $EW_{rest}=6.1\pm0.4 $ \AA, which makes marginal its inclusion in the BL Lacs class. To evaluate a more physical discriminant parameter,  the optical beaming factor $\delta$, as discussed by \cite{Farina2012} and \cite{Landoni2013}, was calculated. This parameter quantifies the contribution  of thermal disk to the total luminosity \citep{Decarli2011}. A $\delta=6.0\pm3.6$ places the source as a BL Lac  just above  the intermediate region between pure QSO ($\delta \simeq1$) and BL Lacs ($\delta\gtrsim$4)  \citep{Landoni2013}.
\\

\textbf{J135120.84+111453.0}: An absorption feature at $\lambda=4530$ \AA\   is detected. If the interpretation is in terms of an Mg II $\lambda2798$ intervening system, a redshift  lower limit is 0.619.  This absorption feature at  $\lambda=4530$  \AA\  was recently  observed also
by \cite{Shaw2013}.
\\

\textbf{J144052.93+061016.1}:  Our  spectrum exhibits Ca II H/K  $\lambda\lambda3934, 3968$   and G Band $\lambda4305$  absorptions lines from underlying host galaxy at $z  =  0.396$. 
Two optical spectra are gathered by SDSS on MJD 53494 and  55686 and  inferred redshifts are given as  unreliable. \cite{Shaw2013} detected an absorption feature  $\lambda\sim 3685$ \AA, that  interpreted as Mg II set a lower limit at 0.316.
\\

 \textbf{J163716.73+131438.7}:  In our spectrum we detect a narrow emission line at  $\lambda= 6170$ \AA. This is most probably a real feature, since it clearly appears on each of the three individual spectra. It can be identified  with O II  $\lambda2737$ at $z = 0.656$. At  the same redshift two absorption lines ascribed to  Ca II H/K $\lambda\lambda3934, 3968$  are apparent. In addition, the position of an absorption at $\lambda=7124$ \AA, encompassed by two H$_{2}$O telluric bands, is consistent with G band absorption at the same redshift of the other lines. Our redshift  is one of the highest ever been detected in optical range by host galaxy absorption lines.
\\

 \textbf{J214406.27-002858.1}:   Because of  the featureless spectrum, a redshift  lower limit of $\sim$ 0.34 is derived  from  $EW_{min}  = 1.36$ \AA. 
 Unreliable redshifts were assigned by the SDSS. 
\\

\textbf{J224448.09-000619.3}:  We clearly detected Ca II H/K $\lambda\lambda3934, 3968$  absorptions  at redshift  $z = 0.641$. The position of an absorption at $\lambda=7057$  \AA\  in a region inside the H$_{2}$O telluric bands, is consistent with G band $\lambda4305$ absorption at the same redshift of Ca II doublet wavelengths. No reliable redshift was obtained by the SDSS.
\\

\textbf{J224730.18+000006.4}: We distinctly observed an absorption feature at $\lambda=5311$ \AA\  with an $EW= +3.0\pm 0.1$\ \AA, that was interpreted as Mg II $\lambda2798$ intervening system setting $z > 0.898$. There is no evidence of the emission line feature detected by \cite{Shaw2013} at  $\lambda \sim 5460$\  \AA\  and taken as Mg II at $z = 0.949$. We suspect this feature to be  instrumental   since it appears in other spectra reported by Shaw with the same spectrograph.

\section{High Energy Emission of the Sources}

The \textit{Fermi} Second Source Catalog \citep[2LAC, ][]{Nolan2012} lists the 1873 significant sources detected by the Large Area Telescope \citep[LAT,][]{Atwood2009} during \textit{Fermi}'s first two years of sky survey observations. Most of  them are jet dominated  AGN.  Among them, more than 400 \textit{Fermi} BL Lacs  attest about their large contribution to  the $\gamma$ emission background among  the brightest extra-Galactic  sources.
To fully describe our small sample of optically selected BL Lacs,  the detections of the target objects  at high and very high energy were investigated. 
 A comparison with TevCat\footnote{\texttt{http://tevcat.uchicago.edu/} } \citep{Horan2008} indicates that there are no TeV counterparts. 
 This is not surprising, since measured redshifts or a lower limits to the redshift are beyond the extragalactic background light horizon, with the exception of J144052.93+061016.1.
 
We have cross-correlated  the \textit{Fermi} archived events available on line\footnote{\tt http://fermi.gsfc.nasa.gov/ssc/data/access/lat/2yr\_catalog/}\ with the positions of our sources to update with respect to LAT 2 release. As shown in Table \ref{fermi}, four of our sources enter in the 2LAC catalogue  \citep[LAT AGN Catalog, ][]{Ackermann2011}. 
Then, we have analyzed all the available 57 month survey data, from the start of the  \textit{Fermi} activity on 2008 August 4 (MJD 54682) to  2013 April 8 (MJD 56390), with the aim of updating the values of flux and photon index, tracing the light curves and looking for new detections. We used the {\tt LAT Science Tools v. 9.27.1}, the Instrument Response Function (IRF) {\tt P7SOURCE\_V6}, the corresponding background files,  following standard procedures\footnote{{\tt http://fermi.gsfc.nasa.gov/ssc/data/analysis/scitools/}}. 
 We selected for each source all the events of class 2 ("source'' type) included in a circular region centered on the optical coordinates and with radius $10^{\circ}$. 
The final source list was determined, by applying a significance threshold.

Two new $\gamma$-ray sources appear: J125032.58+021632.1 and J163716.73+131438.8. Some targets, even if included in 2LAC,  have not been detected on the basis on their $\gamma$ fluxes  over the entire 57 months period, but due to the variability they are found on monthly scale. In Table \ref{fermi} the integrated photon flux in 0.1-100 GeV range or upper limits, the photon index and Test Statistic \citep[TS,][]{Mattox1996} are given  for the whole time of observation in the central columns, while  the right columns refer to monthly detections with highest TS. We considered as valid the results of the likelihood with $TS\geqslant9$, corresponding to about $3\sigma$.

In order to compare these results with the whole dataset, we have  correlated  the  list of 637 radio loud optically selected BL Lacs of P10  with 2LAC, finding 125 positional coincidences,  corresponding to $\sim$ 20 \% of the objects. In the sample examined here we observed a higher percentage, possibly as a consequence  of  the imposed magnitude limit, and the choice of SDSS lineless objects, which can be indicative  of strong beaming.  We have therefore selected the 194 P10 objects  which are lineless and with $r<19.4$, and found 69 correlations with 2LAC, corresponding to $\sim$ 35 \%, which is consistent with our findings.

\section{Summary and Conclusions}

We obtained optical spectroscopy of  a small sample of  BL Lacs  of unknown  redshift. On the basis of the $\gamma$ properties the objects appear representative of the parent sample. In one case a broad line emission was found, in others absorptions from the host galaxy or intervening material were detected. For two objects the spectrum remained featureless, and in these cases the redshift should  be $z\gtrsim0.4$.  
New surveys have  allowed to derive the spectroscopic redshifts of a large number  of BL Lac objects with high S/N spectra. Nevertheless a significant fraction remains of unknown $z$ objects.
The new redshifts are higher compared to the ones of recently assembled large samples (P08, P10, S13). If tentative attributions and lower limits are included, redshift medians for high confidence BL Lacs are 0.39 in P08, 0.43 in P10 radio-loud subsample, 0.32 in S13 and  0.64 for our objects.
Although it is a small sample, it suggests  that a significant fraction 
of most  still unknown $z$ objects  is probably at high $z$ and significantly beamed. 
Larger redshift completeness fraction and  homogeneous and unbiased selections could also better define a deeper picture of the cosmological evolution.
A search for $\gamma$ counterpart emission shows that six out of nine are Fermi $\gamma$-ray emitters with two new detections.
High $z$ and high beamed BL Lacs  deserve new approach and capabilities to derive their redshifts. In the region of   $z\sim 0.1-0.7 $ a very effective  technique was introduced  in the far-UV  ($1135-1800$ \AA) with HST+COS \citep[e.g.][]{Stocke2011, Danforth2010, Danforth2013}, for constraining quite stringently $z$ by  intervening intergalactic medium absorbers detected in $Ly\alpha$, and in $Ly\beta$ and/or metal
lines. An interesting  possibility  is derive $z$ lower limits  for  BL Lacs objects  at redshift $z>1.5$ by searching  for weak and narrow  $Ly{\alpha}$ absorption in optical range from the ground similarly as performed and adopted in UV spectra. Good candidates can be found also in the sample presented here. High resolution spectroscopy is required combined with large diameter telescopes.
 
\begin{table*}
\begin{center}
\caption{Journal of observations.\label{journal} }	
\scriptsize
\begin{tabular}{lllcccclr}
&&&&&&&&\\
&&&&&&&&\\
\tableline
&&&&&&&&\\
  Source				&R.A.		&DEC. 				&Date		&r			&Exp.Time&N	&Seeing	&S/N\\
		         			&  [h:m:s] 		& [d:m:s] 				&			&[mag]		&[min]	&	&[arcsec]	&	\\
(a)		        			& (b)      		&(c)					&(d)			&(e)			& (f)		&(g)	&	(h)	&(i)\\
&&&&&&&&\\
\tableline\tableline
&&&&&&&&\\
J003808.50+001336.5   	&00 38 08.503 	&+ 00 13 36.53 		&2009-06-13	&19.30		&90		&6	&0.6		&50\\
J125032.58+021632.1	&12 50 32.581	&+ 02 16 32.173		&2009-04-30 	&19.22		&45		&3	&0.9		&30\\ 
J135120.84+111453.0 	&13 51 20.847 	&+ 11 14 53.02 		&2009-06-24	&18.58		&45		&3	&1.0		&100\\
J144052.93+061016.1 	&14 40 52.94 	&+ 06 10 16.2			&2009-06-24  	&17.17		&45		&3	&1.1		&140\\
J145507.44+025040.2 	&14 55 07.443	&+ 02 50 40.25 		&2009-08-12 	&19.40		&45		&3	&0.7		&35\\
J163716.73+131438.8 	&16 37 16.737 	&+ 13 14 38.80			&2009-04-29	&18.95		&45		&3	&1.2		& 90\\
J214406.27-002858.1	&21 44 06.271 &$-$ 00 28 58.19 		&2009-05-19	&19.20		&45		&3	&0.7 		&15\\
J224448.09-000619.3 	&22 44 48.095 &$-$ 00 06 19.49		&2009-08-15 	&19.11		&45		&3	&0.5		&90\\
J224730.19+000006.4 	&22 47 30.196	&+ 00 00 06.463 		&2009-09-09	&18.26		&45		&3	&1.1		&45\\
&&&&&&&&\\
\tableline
\end{tabular}
\end{center}
\tablecomments
{(a) Object ID. (b),(c) ICRS Right ascension and Declination Coordinates (J2000).  (d) Date of observation. (e) r apparent PSF magnitude from SDSS DR7. (f) Total exposure time. (g) Number of collected spectra. (h) Seeing during the observation. (i) Signal to noise ratio evaluated as the average over the whole spectrum range, avoiding the regions affected by emission or absorption features.}
\end{table*}

\begin{table*}
\begin{center}
\caption{Spectral Line Parameters \label{par}}
\scriptsize
\begin{tabular}{lcllllllll}
\tableline\tableline
Source				&z			&$\alpha_{\nu}$ &$EW_{min}$	&Line ID		&$\lambda_{\ line}$	&Type	&FWHM		&$EW$				\\				
					&			&			&[\AA]		&			&	[\AA]				&		&[km/s] 		&[\AA]						\\	
	(a)				&(b)			&	(c)		&(d)			&	(e)		&(f)					&(g)		&(h) 			&(i)						\\	
\tableline
&&&&&&&&&\\

J003808.50+001336.5	&$ 0.708<z\lesssim\ 2.7$	&1.14		&0.78		&Mg II		&4780			&a		&\ 900 $\pm$ 200		&\  $+$1.6 $\pm$ 0.3	\\

&&&&&&&&&\\
J125032.58+021632.1	&0.955		&1.44		&1.05		&Mg II		&5469			&e		&4500 $\pm$ 200		&$-$12.1$ \pm$ 0.8		\\	
					&			&			&			&O II			&7288			&e		& \ 700 $\pm$ 200		&\ $-$3.6 $\pm$ 0.8				\\

&&&&&&&&\\	
J135120.84+111453.0 	&$ 0.619<z\lesssim\  2.4$ 	&0.73		&0.31		&Mg 	II 		&4530			&a		&1500 $\pm$ 300		&\ $+$1.0 $\pm$ 0.2				\\

&&&&&&&&\\
J144052.93+061016.1 	&0.396 		&1.08		&0.28		&Ca II		&5491			&g		&1500 $\pm$ 600		&\ +0.4 $\pm $ 0.1					\\	
					&			&			&			&Ca II		&5542			&g		&1000 $\pm$ 200		&\ +0.4 $\pm$ 0.1					\\	
					&			&			&			&G band		&6008			&g		&1750 $\pm$ 50 		& +0.93 $\pm$ 0.04				\\

&&&&&&&&&\\
J145507.44+025040.3	&$0.47<z\lesssim\ 2.5$		&1.44		&0.79		&featureless	&.....					&.....		&..........					&..........							\\					

&&&&&&&&&\\	
J163716.73+131438.7	&0.655		&$$1.19		&0.40		&O II			&6170			&e		&900$\pm$100 	&$-$0.8$\pm$0.2		\\ 
					&			&			&			&Ca II		&6523			&g		&1100$\pm$100	&+0.7$\pm$0.07		\\
					&			&			&			&Ca II		&6566			&g		&800$\pm$300		&+0.4$\pm$0.09		\\
					&			&			&			&G band 		&7124			&g		&800$\pm$200		&+0.8$\pm$0.2		\\

&&&&&&&&&\\

J214406.27$-$002858.1 	&$ 0.34<z\lesssim\ 2.5$		&1.36		&1.45		&featureless	&.....				&.....		&..........					&..........						\\

&&&&&&&&&\\
J224448.09$-$000619.3 	&0.640 		&0.88		&0.35		&Ca II		&6450			&g		&1300 $\pm$ 100		&+0.75 $\pm$ 0.05				\\
					&			&			&			&Ca II		&6511		&g		&1300 $\pm$ 300		&\ +0.7 $\pm$ 0.1					\\
					&			&			&			&G band		&7057			&g		&1200 $\pm$ 300		&+0.55 $\pm$ 0.07				\\

&&&&&&&&\\

J224730.18+000006.4	&$0.898<z\lesssim\ 2.5$ 	&1.27		&0.52		&Mg II 		&5311			&a		&1800 $\pm$ 100 		&\ +3.0 $\pm$ 0.1			\\

&&&&&&&&\\
 \tableline
\end{tabular}
 \end{center}
\tablecomments{(a) Object ID. (b) Average redshift from the single lines or limits: lower limits  from intervening systems  or following \cite{Sbarufatti2005b} for featureless spectra; upper limits from absence of  $Ly\alpha$ absorptions following \cite{Shaw2013} and references therein. (c) Spectral index of the continuum, defined by $F_{\nu}\propto\nu^{-\alpha_{\nu}}$. (d) Minimum detectable equivalent width  (observer frame). (e) Line identification. (f) Wavelength at the center of the feature. (g) Type of feature: e: emission line; g: host galaxy absorption line; a: intervening system absorption line. (h) Full Width Half Maximum of the line. (i)  Line equivalent width (observer frame).}
 \end{table*}

\clearpage
\begin{landscape}
\begin{table*}
\begin{center}
\caption{Fermi LAT detections for the target objects. \label{fermi}}
\tiny 
\begin{tabular}{cccccc|ccc|cccc}
\tableline\tableline
&&&&&&&&&&\\ 
Source				&in	&2 FGL 			&R.A.	&Dec.	&error 		&$F_{0.1-100 \ GeV}	$		& Photon				&TS		&Time	 	&$f_{0.1-100 \ GeV}$		&Photon	&$TS_h$\\
					&2LAC&name			&[deg]	&[deg]	&[deg]		&x10$^{-9}$				&Index				&		&			&x10$^{-8}$				&Index	&	\\					
					&	&				&		&		&			&[ph\ cm$^{-2}$ s$^{-1}$] 		& 					&		&			&[ph\ cm$^{-2}$ s$^{-1}$]		&		&\\ 
	(a)				&	&		(b)		&(c)		&(c)		&(d)			&(e)						&	(f)				&(g)		&	(h)		&(i)						&	(j)	&(k)	\\
&&&&&&&&&&&&\\ 
\hline
&&&&&&&&&&&&\\ 
J003808.50+001336.5	&Y&J0038.1+0015		&9.542	&0.265	&0.217		&1.6$\pm$0.7				&$-$1.78$\pm$0.15		&42		&2009/10/09	&1.80$\pm$0.16			&1.74$\pm$0.04	&24\\
J125032.58+021632.1	&....	&	....			&192.578	&2.308	 &0.095		&2.2$\pm$1.1				&$-$1.85$\pm$0.17		&35		&2009/11/08	&2.07$\pm$0.54			&1.93$\pm$0.13	&15\\
J135120.84+111453.0     &Y&J1351.4+1115		&207.867	&11.256	&0.110		&$<$ 2 (5$\sigma$)			&....					&....		&2012/04/26	&1.16$\pm$0.84			&1.63$\pm$0.30	&24\\
J144052.93+061016.1	&Y&J1440.9+0611		&220.248	&6.189	&0.099		&9.6$\pm$2.1	 			&2.1$\pm$0.1			&148	&2008/12/13	&2.72$\pm$\ 1.3			&1.93$\pm$0.23	&34\\
J145507.44+025040.2	&....&		....			&....		&....		&....			&$<$ 2 (5$\sigma$)			&....					&....		&....			&....						&....				&	\\
J163716.73+131438.8	&....&		....			&249.37	&13.20	&0.12		&$<$ 2 (5$\sigma$)			&....					&....		&2011/09/29	&10.2$\pm$\ 4.4			&3.69$\pm$0.60	& 9\\
J214406.27-002858.1	&....&		....			&....		&....		&....			&$<$ 2 (5$\sigma$)			&....					&....		&....			&....						&....				&\\	
J224448.09-000619.3 	&....&		....			&....		&....		&....			&$<$ 2 (5$\sigma$)			&....					&....		&....			&....						&....				&\\	
J224730.19+000006.4	&Y&J2247.2$-$0002	&341.811	&-0.049	&0.152		&$<$ 2 (5$\sigma$)			&....					&....		&2012/07/25	&1.99$\pm$\ 1.3			&1.93 $\pm$0.30	&16\\

&&&&&&&&&&&&\\ 	
 \hline
\end{tabular}
 \end{center}
\tablecomments{(a) Object ID. (b) Fermi Gamma-ray LAT designation in 2-year catalog. (c) $\gamma$ counterpart coordinates. (d) $95\%$ error radius. (e): Integral photon flux in 0.1-100 GeV range. (f) Photon Index defined as $\nu F_{\nu}\propto \nu^{-\Gamma+2}$.  (g) Test Statistic \citep{Mattox1996}. (h) Time of highest significance observation: measures derived from 30 days integration around the date ($\pm$15 d) in column (h).  (i) Highest significance photon flux. (j) Photon Index of highest significance observation. (k) Highest Test Statistic.}.
 \end{table*}
\end{landscape}

 \setcounter{figure}{0}
   \begin{figure*}
          \centering
            \caption{Spectra of the objects in the observer frame. \textit{Top panel: }Flux density spectrum in Relative Units (R. U.);  \textit{Bottom panel:}  Slightly smoothed normalized spectrum (3 pixel kernel). Telluric  bands are flagged by  $\oplus$, spectral lines are marked  by line identifications, absorption features from interstellar medium of our galaxy are labeled by ISM, diffuse interstellar bands by DIB.}
         \label{4_1}
            \includegraphics[width=19cm]{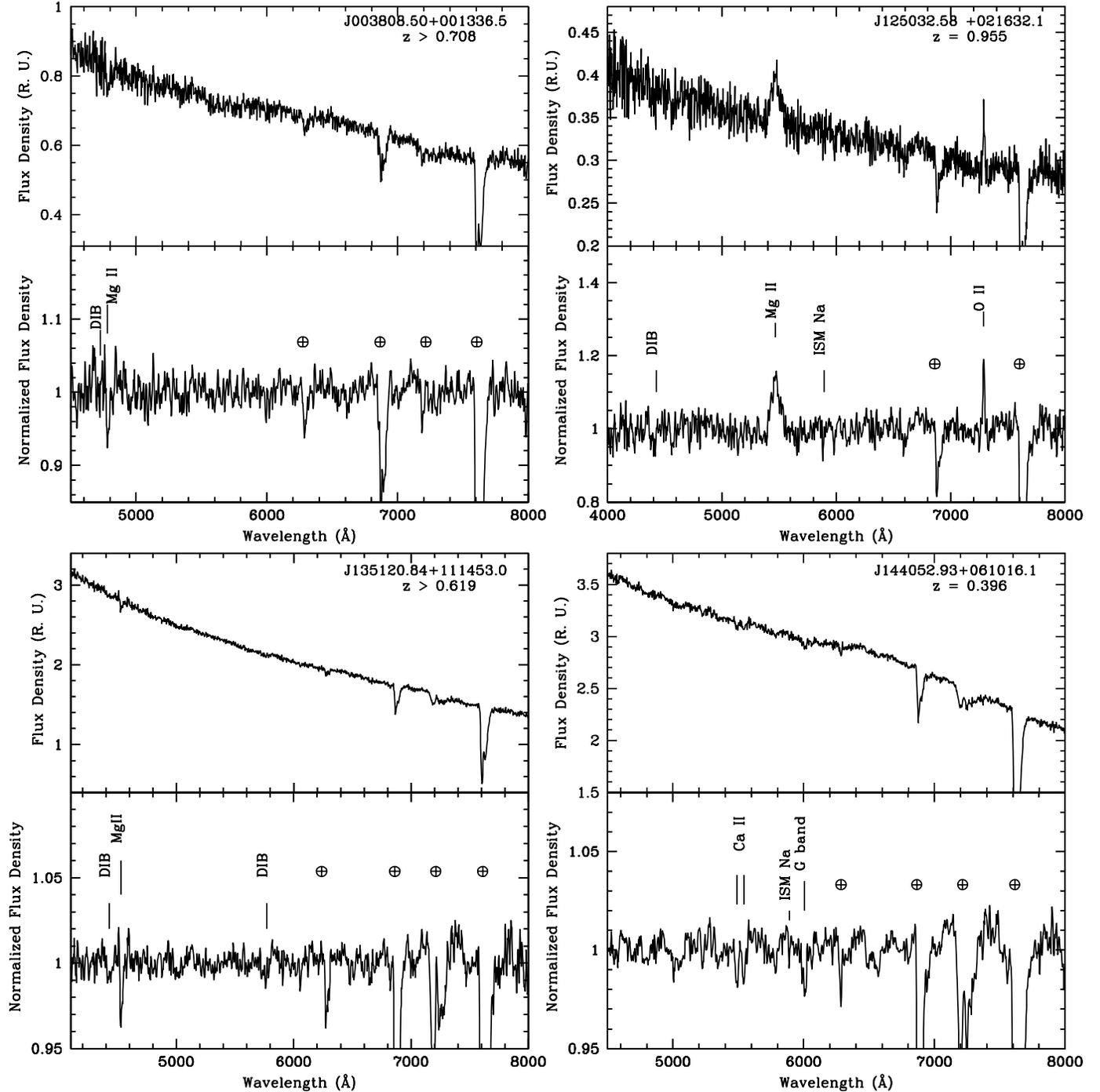}            
   \end{figure*}   
    \setcounter{figure}{0}
  \begin{figure*}
          \centering
            \caption{--- Continued}
            \includegraphics[width=19cm]{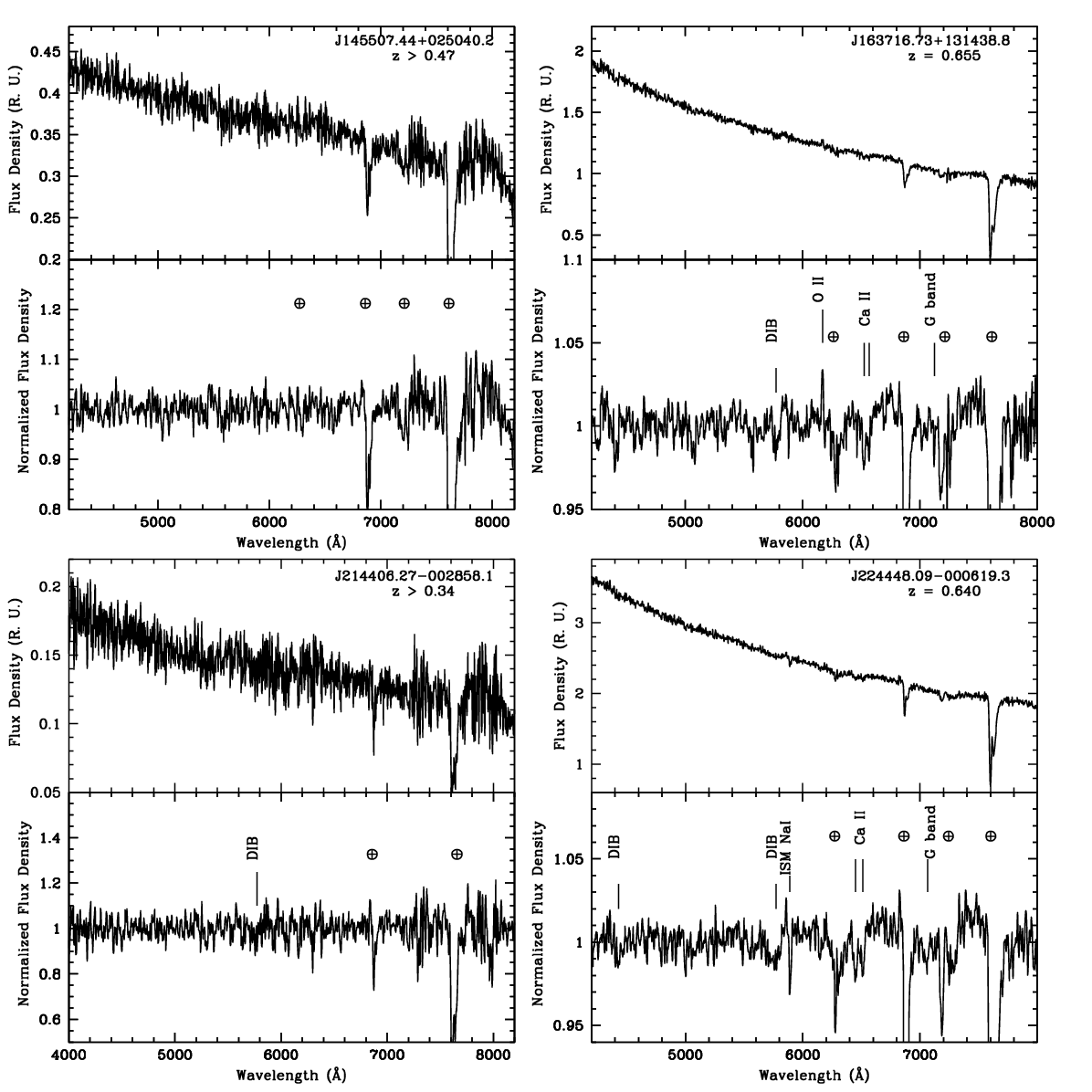}            
    \end{figure*}
    
     \setcounter{figure}{0}
      \begin{figure*}
          \centering
            \caption{--- Continued}
            \includegraphics[width=19cm]{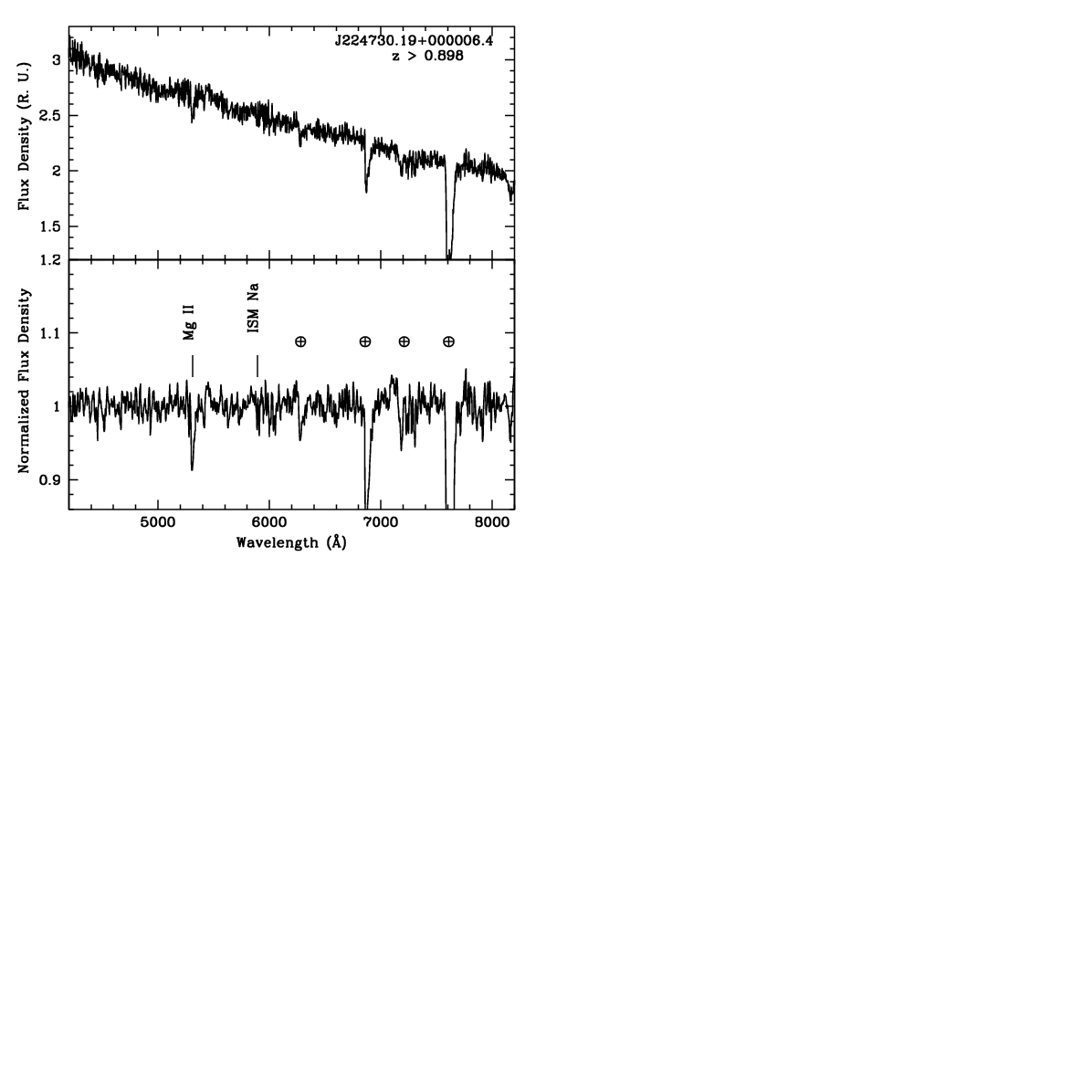}            
    \end{figure*}


\begin{thebibliography}{}

\bibitem[Abazajian et al.(2009)]{Abazajian2009} Abazajian, K.~N., 
Adelman-McCarthy, J.~K., Ag{\"u}eros, M.~A., et al.\ 2009, \apjs, 182, 543 

\bibitem[Ackermann et al.(2011)]{Ackermann2011} Ackermann, M., 
Ajello, M., Allafort, A., et al.\ 2011, \apj, 743, 171 

\bibitem[Ajello et al.(2009)]{Ajello2009} Ajello, M., Costamante, 
L., Sambruna, R.~M., et al.\ 2009, \apj, 699, 603 

\bibitem[Atwood et al.(2009)]{Atwood2009} Atwood, W.~B., Abdo, 
A.~A., Ackermann, M., et al.\ 2009, \apj, 697, 1071 

\bibitem[Bade et 
al.(1998)]{Bade1998} Bade, N., Beckmann, V., Douglas, N.~G., et al.\ 1998, \aap, 334, 459 

\bibitem[Becker et al.(1995)]{Becker1995} Becker, R.~H., White, 
R.~L., \& Helfand, D.~J.\ 1995, \apj, 450, 559

\bibitem[Beckmann et 
al.(2003)]{Beckmann2003} Beckmann, V., Engels, D., Bade, N., \& Wucknitz, O.\ 2003, \aap, 401, 927

\bibitem[Blandford 
\& Rees(1978)]{Blandford1978} Blandford, R.~D., \& Rees, M.~J.\ 1978, \physscr, 17, 265 

\bibitem[Caccianiga et al.(2002)]{Caccianiga2002} Caccianiga, A., 
Maccacaro, T., Wolter, A., Della Ceca, R., 
\& Gioia, I.~M.\ 2002, \apj, 566, 181 

\bibitem[Carangelo et al.(2003)]{Carangelo2003} Carangelo, N., 
Falomo, R., Kotilainen, J., Treves, A., 
\& Ulrich, M.-H.\ 2003, High Energy Blazar Astronomy, 299, 299 

\bibitem[\protect\citeauthoryear{Cardelli, Clayton, \& Mathis}{1989}]{Cardelli1989} Cardelli J.~A., Clayton G.~C., Mathis J.~S., 1989, ApJ, 345, 245 

\bibitem[Collinge et al.(2005)] {Collinge2005} Collinge, M.~J., 
Strauss, M.~A., Hall, P.~B., et al.\ 2005, \aj, 129, 2542 

\bibitem[Condon et al.(1998)]{Condon1998} Condon, J.~J., Cotton, 
W.~D., Greisen, E.~W., et al.\ 1998, \aj, 115, 1693 

\bibitem[Croom et al.(2004)]{Croom2004} Croom, S.~M., Smith, 
R.~J., Boyle, B.~J., et al.\ 2004, \mnras, 349, 1397


\bibitem[Danforth et al.(2010)]{Danforth2010} Danforth, C.~W., 
Keeney, B.~A., Stocke, J.~T., Shull, J.~M., 
\& Yao, Y.\ 2010, \apj, 720, 976

\bibitem[Danforth et al.(2013)]{Danforth2013} Danforth, C.~W., 
Nalewajko, K., France, K., \& Keeney, B.~A.\ 2013, \apj, 764, 57

\bibitem[Decarli et al.(2011)]{Decarli2011} Decarli, R., Dotti, M., 
\& Treves, A.\ 2011, \mnras, 413, 39 

\bibitem[Falomo et al.(1987a)]{Falomo1987a} Falomo, R., Maraschi, 
L., Treves, A., 
\& Tanzi, E.~G.\ 1987,a, Liege International Astrophysical Colloquia, 27, 153

\bibitem[Falomo et al.(1987b)]{Falomo1987b} Falomo, R., Maraschi, 
L., Treves, A., \& Tanzi, E.~G.\ 1987,b, \apjl, 318, L39


\bibitem[Falomo(1990)]{Falomo1990} Falomo, R.\ 1990, \apj, 353, 
112005
 
\bibitem[Farina et al.(2012)]{Farina2012} Farina, E.~P., Decarli, 
R., Falomo, R., Treves, A., \& Raiteri, C.~M.\ 2012, \mnras, 424, 393

\bibitem[Giommi et al.(1999)]{Giommi1999} Giommi, P., Menna, 
M.~T., \& Padovani, P.\ 1999, \mnras, 310, 465

\bibitem[Giommi et 
al.(2009)]{Giommi2009} Giommi, P., Colafrancesco, S., Padovani, P., et al.\ 2009, \aap, 508, 107

\bibitem[Giommi et al.(2012)]{Giommi2012} Giommi, P., Padovani, 
P., Polenta, G., et al.\ 2012, \mnras, 420, 2899


\bibitem[Green et al.(1986)]{1986ApJS...61..305G} Green, R.~F., Schmidt, 
M., \& Liebert, J.\ 1986, \apjs, 61, 305 

\bibitem[Horan 
\& Wakely(2008)]{Horan2008} Horan, D., \& Wakely, S.\ 2008, AAS/High Energy Astrophysics Division \#10, 10, \#41.06 

\bibitem[Landoni et al.(2013)]{Landoni2013} Landoni, M., Falomo, 
R., Treves, A., et al.\ 2013, \aj, 145, 114 

\bibitem[Kimball 
\& Ivezic(2007)]{Kimball2007} Kimball, A.~E., \& Ivezic, Z.\ 2007, Bulletin of the American Astronomical Society, 39, \#132.19 

\bibitem[Marcha et al.(1996)]{Marcha1996} Marcha, M.~J.~M., 
Browne, I.~W.~A., Impey, C.~D., \& Smith, P.~S.\ 1996, \mnras, 281, 42

\bibitem[Mattox et al.(1996)]{Mattox1996} Mattox, J.~R., Bertsch, 
D.~L., Chiang, J., et al.\ 1996, \apj, 461, 396 

\bibitem[Nolan et al.(2012)]{Nolan2012} Nolan, P.~L., Abdo, 
A.~A., Ackermann, M., et al.\ 2012, VizieR Online Data Catalog, 219, 90031 

\bibitem[Padovani 
\& Giommi(1995)]{Padovani1995} Padovani, P., \& Giommi, P.\ 1995, \mnras, 277, 1477

\bibitem[Padovani et al.(2007)]{Padovani2007} Padovani, P., Giommi, 
P., Landt, H., \& Perlman, E.~S.\ 2007, \apj, 662, 182

\bibitem[Plotkin et al.(2008)]{Plotkin2008} Plotkin, R.~M., 
Anderson, S.~F., Hall, P.~B., et al.\ 2008, \aj, 135, 2453 

\bibitem[Plotkin et al.(2010a)]{Plotkin2010a} Plotkin, R.~M., 
Anderson, S.~F., Brandt, W.~N., et al.\ 2010a, \aj, 139, 390

\bibitem[Plotkin et al.(2010b)]{Plotkin2010b} Plotkin, R.~M., 
Anderson, S.~F., Brandt, W.~N., et al.\ 2010b, \apj, 721, 562 

 \bibitem[Rector et al.(2000)]{Rector2000} Rector, T.~A., Stocke, 
J.~T., Perlman, E.~S., Morris, S.~L., \& Gioia, I.~M.\ 2000, \aj, 120, 1626 

\bibitem[Rector 
\& Stocke(2001)]{Rector2001} Rector, T.~A., \& Stocke, J.~T.\ 2001, \aj, 122, 565 

\bibitem[Sbarufatti et al.(2005a)]{Sbarufatti2005a} Sbarufatti, B., 
Treves, A., \& Falomo, R.\ 2005,a, \apj, 635, 173 

\bibitem[Sbarufatti et al.(2005b)]{Sbarufatti2005b} Sbarufatti, B., 
Treves, A., Falomo, R., et al.\ 2005,b, \aj, 129, 559  Paper I

\bibitem[Sbarufatti et al.(2006a)]{Sbarufatti2006a} Sbarufatti, B., 
Treves, A., Falomo, R., et al.\ 2006,a, \aj, 132, 1 Paper II

\bibitem[Sbarufatti et 
al.(2006b)]{Sbarufatti2006b} Sbarufatti, B., Falomo, R., Treves, A., \& Kotilainen, J.\ 2006,b, \aap, 457, 35 

\bibitem[Sbarufatti et al.(2009)]{Sbarufatti2009} Sbarufatti, B., 
Ciprini, S., Kotilainen, J., et al.\ 2009, \aj, 137, 337

\bibitem[Shaw et al.(2013)]{Shaw2013} Shaw, M.~S., Romani, 
R.~W., Cotter, G., et al.\ 2013, \apj, 764, 135

\bibitem[\protect\citeauthoryear{Schlegel, Finkbeiner, \& Davis}{1998}]{Schlegel1998} Schlegel D.~J., Finkbeiner D.~P., Davis M., 1998, ApJ, 500, 525 

\bibitem[Stickel et al.(1991)]{Stickel1991} Stickel, M., Padovani, 
P., Urry, C.~M., Fried, J.~W., \& Kuehr, H.\ 1991, \apj, 374, 431 

\bibitem[Stickel et 
al.(1993)]{Stickel1993} Stickel, M., Fried, J.~W., \& Kuehr, H.\ 1993, \aaps, 98, 393

\bibitem[Stocke et al.(1985)]{Stocke1985} Stocke, J.~T., Liebert, 
J., Schmidt, G., et al.\ 1985, \apj, 298, 619 

\bibitem[Stocke et al.(1991)]{Stocke1991} Stocke, J.~T., Morris, 
S.~L., Gioia, I.~M., et al.\ 1991, \apjs, 76, 813 

\bibitem[Stocke et al.(2011)]{Stocke2011} Stocke, J.~T., Danforth, 
C.~W., \& Perlman, E.~S.\ 2011, \apj, 732, 113 

\bibitem[\protect\citeauthoryear{Tody}{1986}]{Tody1986} Tody D., 1986, SPIE, 627, 733 

\bibitem[Veron(1994)]{Veron1994} Veron, P.\ 1994, \aap, 283, 802 

\bibitem[Voges et 
al.(1999)]{Voges1999} Voges, W., Aschenbach, B., Boller, T., et al.\ 1999, \aap, 349, 389 

\bibitem[Ulrich(1989)]{Ulrich1989}Ulrich M.H., 1989, in  Maraschi, L., 
Maccacaro, T., \& Ulrich, M.-H.\ 1989, BL Lac Objects, 334,

\bibitem[Wu et al.(2012)]{Wu2012} Wu, J., Brandt, W.~N., 
Anderson, S.~F., et al.\ 2012, \apj, 747, 10

\end{thebibliography}
\end{document}